\documentclass[class=preprint]{nature}
\usepackage{graphicx}
\graphicspath{{Figures/}}

\usepackage[usenames,dvipsnames]{color}
\usepackage{hyperref}
\hypersetup{colorlinks=true,allcolors=blue}
\usepackage{authblk}
\usepackage[a4paper]{geometry}
\usepackage{fixltx2e}
\usepackage{upgreek}
\usepackage{amsmath}
\usepackage{amssymb}
\usepackage{amsthm}
\usepackage{mathrsfs}
\usepackage{amsbsy}
\usepackage{amstext}
\usepackage{xcolor}
\usepackage{float}
\usepackage{multirow}
\usepackage{wasysym}
\usepackage{mathtools}
\usepackage{subfigure}
\usepackage[version=4]{mhchem}
\usepackage{amsfonts}
\usepackage{dcolumn}
\usepackage{placeins}
\usepackage{float}
\usepackage{caption}
\usepackage[right]{lineno}
\usepackage{fancyhdr}
\usepackage{bbold,bm}
\usepackage{pdfpages}
\usepackage[utf8]{inputenc}
\usepackage[T1]{fontenc}%
\usepackage{siunitx}  
\providecommand{\U}[1]{\protect\rule{.1in}{.1in}}
\DeclareCaptionLabelSeparator{vline}{$\boldsymbol{:}$}
\makeatletter
\let\saved@includegraphics\includegraphics
\AtBeginDocument{\let\includegraphics\saved@includegraphics}
\renewenvironment*{figure}{\@float{figure}}{\end@float}
\makeatother

\title{Keldysh tuning of photoluminescence in a lead halide perovskite crystal}

\author[1,*]{Zhuquan Zhang}
\author[2,*]{Honglie Ning}
\author[1,*]{Zi-Jie Liu}
\author[3]{Jin Hou}
\author[3,4]{Aditya D. Mohite}
\author[5]{Edoardo Baldini}
\author[2]{Nuh Gedik}
\author[1,$\dagger$]{Keith A. Nelson}

\affil[1]{Department of Chemistry, Massachusetts Institute of Technology, Cambridge, Massachusetts, 02139, USA.}
\affil[2]{Department of Physics, Massachusetts Institute of Technology, Cambridge, Massachusetts, 02139, USA.}
\affil[3]{Department of Materials Science and NanoEngineering, Rice University, Houston, Texas, 77005, USA.} 
\affil[4]{Department of Chemical and Biomolecular Engineering, Rice University, Houston, Texas, 77005, USA.}
\affil[5]{Department of Physics, The University of Texas at Austin, Austin, Texas, 78712, USA}
\affil[$\dagger$]{e-mail: kanelson@mit.edu}
\affil[*]{These authors contributed equally to this work.}

\begin{document}
\maketitle

%%%%%%%%%%%%%%%%%%%%%%%%
\newpage
\section*{Abstract}
In 1964, Keldysh laid the groundwork for strong-field physics in atomic, molecular, and solid-state systems by delineating a ubiquitous transition from multiphoton absorption to quantum electron tunneling under intense AC driving forces. While both processes in semiconductors can generate carriers and result in photon emission through electron-hole recombination, the low quantum yields in most materials have hindered direct observation of the Keldysh crossover. Leveraging the large quantum yields of photoluminescence in lead halide perovskites, we show that we can not only induce bright light emission from extreme sub-bandgap light excitation but also distinguish between photon-induced and electric-field-induced processes. Our results are rationalized by the Landau-Dykhne formalism, providing insights into the non-equilibrium dynamics of strong-field light-matter interactions. These findings open new avenues for light upconversion and sub-bandgap photon detection, highlighting the potential of lead halide perovskites in advanced optoelectronic applications.

\newpage

\section*{Main Text}
Over the past few decades, lead halide perovskites have emerged as unconventional semiconductors with exceptional functionalities, finding broad applications ranging from photonics\cite{sutherland2016perovskite} and optoelectronics\cite{kovalenko2017properties,fu2019metal} to electronics\cite{zhao2016organic,younis2021halide}. The large quantum yield of perovskite luminescence is exploited in applications such as multiphoton absorption for photon upconversion\cite{chen2022two,zhou2021perovskites} and electroluminescence in light-emitting diodes\cite{fakharuddin2022perovskite,liu2021electroluminescence}. In the former process, infrared electromagnetic fields facilitate the simultaneous absorption of multiple lower-energy photons, leading to luminescence upon electron-hole recombination. In contrast, the latter process generates luminescence through charge injection and recombination enabled by quantum tunneling under a strong DC electric field. With strong AC fields, a tunneling process akin to those observed with DC fields could emerge, a phenomenon seen across various materials\cite{kruchinin2018colloquium,kuehn2010terahertz,hirori2011extraordinary,Ghimire2011RedshiftField,schultze2014attosecond,lange2014extremely,Mayer2015TunnelingExcitation,pein2017terahertz,shi2022terahertz,li2022keldysh,nishidome2023influence,shinjo2024keldysh}. Therefore, there exists a crossover from photon-induced incoherent light emission to electric-field-induced emission. However, the evolution of a photoluminescence (PL) response from a multiphoton process to quantum electron tunneling remains to be established. Demonstrating this tunability in PL could unlock new optoelectronic applications based on lead halide perovskites.

Here, we investigate the origin of PL in the archetypal lead halide perovskite CsPbBr$_3$ under sub-bandgap radiation. By varying the frequency and amplitude of the driving AC fields, we demonstrate the ability to tune the PL across multiphoton and electron tunneling processes on demand. Our findings, explained by Keldysh theory through the Landau-Dykhne formalism, offer new insights for developing highly efficient photon conversion schemes in semiconductors.

\begin{center}
\begin{figure}[H]
\linespread{1}
   \sbox0{\includegraphics[width=\textwidth]{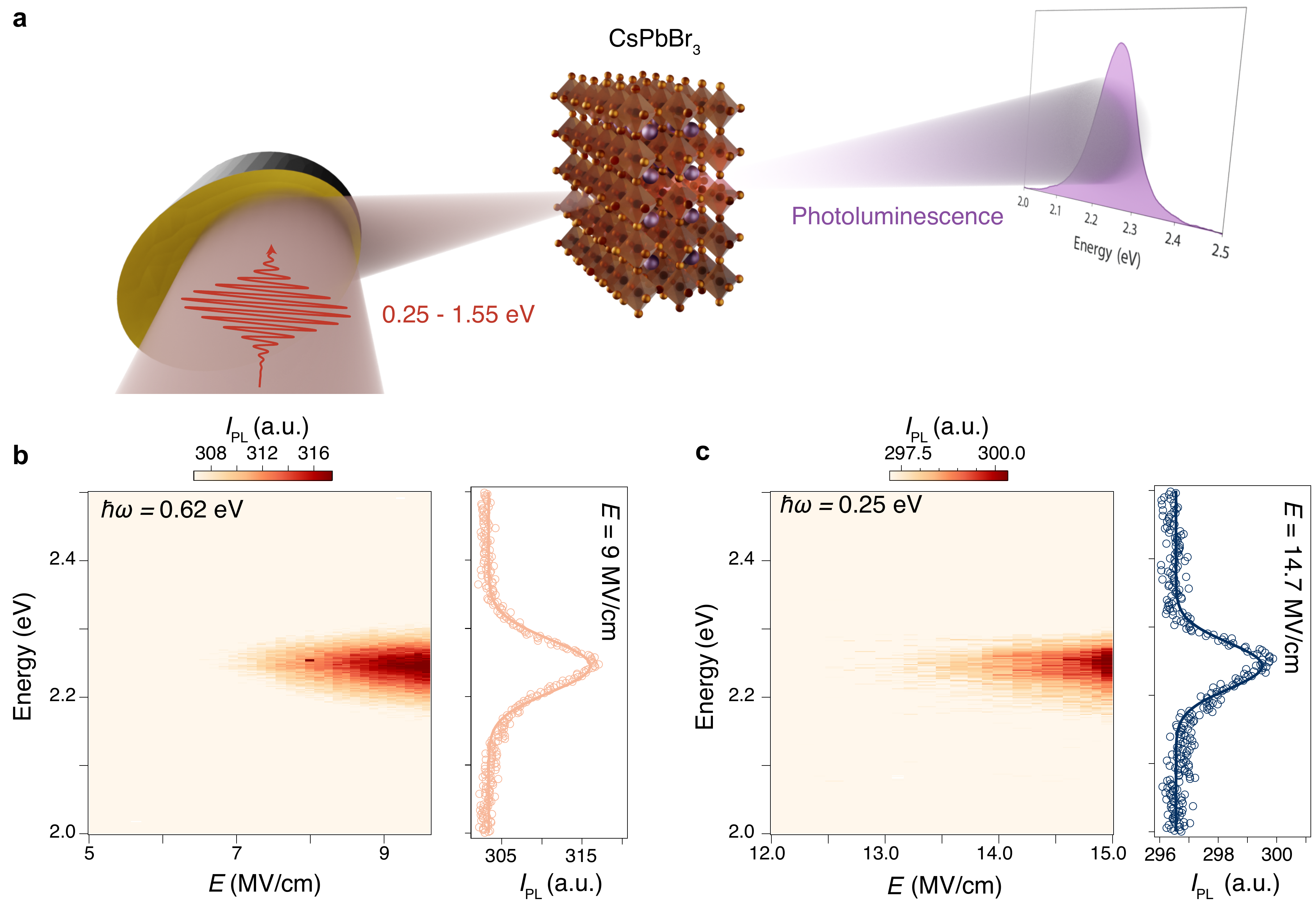}}% measure width
    \begin{minipage}{\wd0}
  \usebox0
  %\captionsetup{justification=raggedright,singlelinecheck=false}
  \captionsetup{labelfont={bf},labelformat={default},labelsep=period,name={Fig.}}
  \caption{\textbf{Experimental scheme and photoluminescence spectra.} \textbf{a,} Schematic illustration of the experimental setup for photoluminescence (PL) measurement. Tailored pump pulses are focused onto CsPbBr$_3$ to generate charge carriers nonlinearly and thereby to induce PL. The resulting PL spectrum is measured as a function of the pump field strength $E$ at each pump photon energy. \textbf{b,} $E$-dependent PL spectra at $\hbar\omega=0.62$ eV and a characteristic spetrum at $E=9$ MV/cm. \textbf{c,} $E$-dependent PL spectra at $\hbar\omega=0.25$ eV and a characteristic spetrum at $E=14.7$ MV/cm. Solid lines are Gaussian fits.} 
  \label{Fig1}
\end{minipage}
\end{figure}  
\end{center}

In our experiment, we generate a tailored infrared laser pulse with photon energy below the band gap of single-crystalline CsPbBr$_3$ and focus it onto the sample at room temperature (see Fig. \ref{Fig1}a). Afterward, we spectrally resolve and measure the resulting PL as a function of the driving electric field amplitude $E$. Figures \ref{Fig1}b and \ref{Fig1}c show the $E$-dependent PL spectra at two selected pump photon energies $\hbar\omega=$ 0.62 and 0.25 eV. From these data, it is evident that prominent PL signals can be detected for both pump photon energies at sufficiently strong pump electric fields. Additionally, a larger PL intensity ($I_{\mathrm{PL}}$) is achieved at $\hbar\omega=$ 0.62 eV even with a smaller pump field strength than that at $\hbar\omega=$ 0.25 eV. This nonlinear carrier generation may stem from multiphoton or quantum electron tunneling processes.

\begin{center}
\begin{figure}[H]
\linespread{1}
   \sbox0{\includegraphics[width=\textwidth]{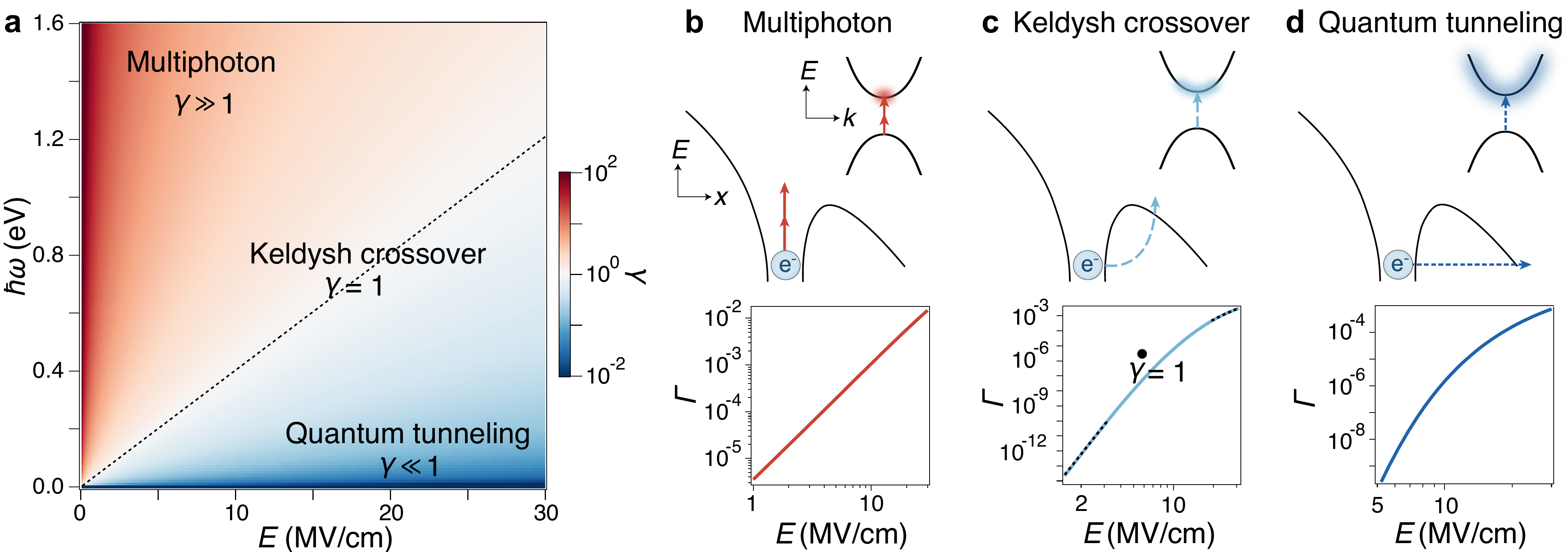}}% measure width
    \begin{minipage}{\wd0}
  \usebox0
  %\captionsetup{justification=raggedright,singlelinecheck=false}
  \captionsetup{labelfont={bf},labelformat={default},labelsep=period,name={Fig.}}
  \caption{\textbf{Keldysh space control.} \textbf{a,} Calculated Keldysh parameter $\gamma$ as a function of the driving field strength $E$ and pump energy $\hbar\omega$. The multiphoton and adiabatic tunneling regions are indicated, with a dashed line denoting $\gamma=1$ where the Keldysh crossover occurs. \textbf{b-d,} Schematic representations of the multiphoton, Keldysh crossover, and adiabatic tunneling processes. Top row: real-space and reciprocal-space diagrams illustrating different regimes of interband transitions under an oscillating electric field. Arrows indicate the excitation pathways of electrons, and the blobs represent the distribution of excited electrons in the conduction band. Bottom row: electron-hole pair production rate as a function of the driving field strength $E$ at selected constant pump energies, plotted on a bi-logarithmic scale. The black dot marks the Keldysh crossover at $\gamma=1$. Dashed lines indicate the scaling relations in the multiphoton and tunneling regimes.} 
  \label{Fig2}
\end{minipage}
\end{figure}  
\end{center}

To distinguish between the two dominant carrier generation processes, Keldysh introduced a theory in 1964 that simultaneously describes both mechanisms for the ionization of atomic gases and crystals\cite{Keldysh1965IonizationWave}. In his seminal paper, Keldysh defined a dimensionless parameter, $\gamma = \frac{\omega \sqrt{m \Delta}}{eE}$, where $m$ is the reduced effective mass of the charge carrier, $e$ is the fundamental quantum charge, and $\Delta$ is the band gap. This quantity, now known as the Keldysh parameter, classifies interband transitions into three regimes: multiphoton processes ($\gamma \gg 1$), diabatic tunneling processes ($\gamma \simeq 1$), and adiabatic tunneling processes ($\gamma \ll 1$). By tuning both the frequency ($\omega$) and the electric field strength ($E$) of the AC pump field, one can achieve precise control over the Keldysh parameter and thereby over interband transitions, as illustrated in Fig. \ref{Fig2}a. In the multiphoton regime ($\gamma \gg 1$, Fig. \ref{Fig2}b), interband transitions can occur through the absorption of integer numbers of photons. Consequently, the electron-hole pair production rate $\Gamma$ scales as a power law, i.e., $\Gamma \propto E^{2a}$, represented by a straight line on a bi-logarithmic scale, with $a$ being the number of photons participating in the multiphoton process. As $\gamma$ approaches unity, the electron-hole pair production rate deviates from the power-law scaling, transitioning to the diabatic tunneling process at the Keldysh crossover (see Fig. \ref{Fig2}c). In this regime, the charge carriers are unable to follow the rapidly oscillating driving fields adiabatically, but the classically forbidden region of the conduction band becomes accessible. For $\gamma \ll 1$, the adiabatic tunneling process dominates, with the AC pump field acting primarily as a classical electric field. Here, the electron-hole pair production rate exhibits threshold behavior, i.e., $\Gamma\propto E\exp(-\pi E_{th}/E)$ (see Fig. \ref{Fig2}d). Therefore, the dependence of $\Gamma$ on $E$ provides distinct signatures to differentiate among these regimes and serves as the benchmark for observing the Keldysh crossover.

\begin{center}
\begin{figure}[H]
\linespread{1}
   \sbox0{\includegraphics[width=\textwidth]{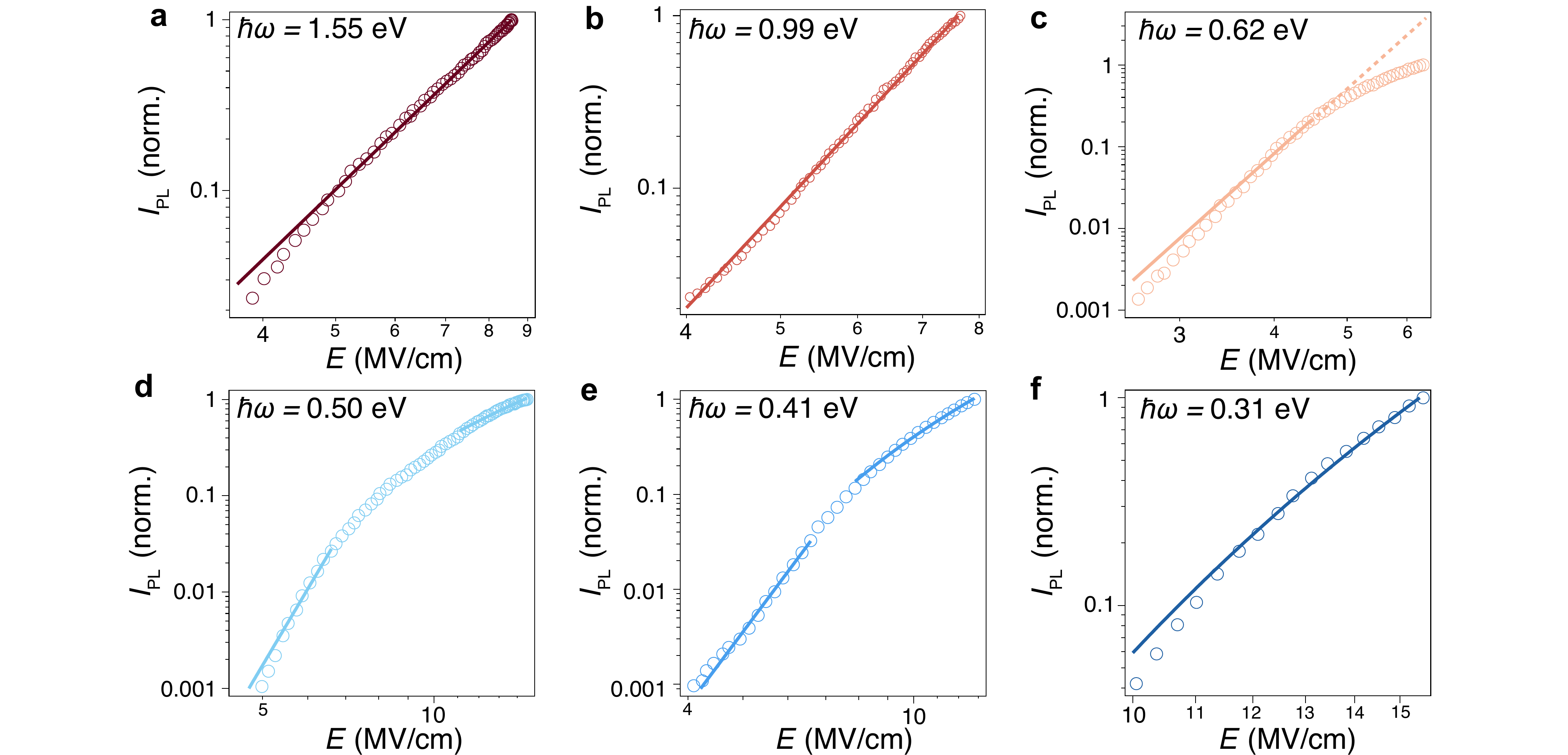}}% measure width
    \begin{minipage}{\wd0}
  \usebox0
  %\captionsetup{justification=raggedright,singlelinecheck=false}
  \captionsetup{labelfont={bf},labelformat={default},labelsep=period,name={Fig.}}
  \caption{\textbf{Experimentally measured $E$-dependent PL intensity.} PL intensity as a function of the driving field strength $E$, normalized by the maximum value at various pump energies $\hbar\omega$. Solid lines in panels \textbf{a}-\textbf{c} and at low $E$ values in panels \textbf{d},\textbf{e} represent multiphoton fits $E^{2a}$ (see Methods). The dashed line in panel \textbf{c} indicates the deviation of the experimental data from the multiphoton fit in the Keldysh crossover regime. Solid lines in panel \textbf{f} and at high $E$ values in panels \textbf{d},\textbf{e} correspond to adiabatic tunneling fits $E\exp(-\pi E_{th}/E)$ (see Methods).} 
  \label{Fig3}
\end{minipage}
\end{figure}  
\end{center}

To experimentally investigate the Keldysh crossover in PL generation, we measure the PL intensity dependence on the pump electric field strength $E$ over a wide range of pump photon energies $\hbar\omega$. Since the peak position and lineshape of the PL remain relatively unaffected by different pump electric fields and pump photon energies, as shown in Fig. \ref{Fig1} (also see Supplementary Note 2), we can spectrally filter and isolate the PL signal using a point detector to enhance detection sensitivity (see Methods). CsPbBr$_3$ serves as an ideal model system due to the absence of excitonic correlations at room temperature and the primarily monomolecular nature of its PL\cite{saouma2017multiphoton,clark2016polarization,abiedh2021mixed}. Thus, $I_{\mathrm{PL}} \propto n$, where $n$ is the number of generated charge carriers. Since the electron-hole pair production rate $\Gamma$ is also proportional to $n$, we have $I_{\mathrm{PL}} \propto \Gamma$. We confirm this relationship by initially pumping the sample at $\hbar\omega=1.55$ eV, as shown in Fig. \ref{Fig3}a. Given the band gap $\Delta$ of approximately 2.3 eV, we expect a two-photon absorption at this pump photon energy. Consequently, if $I_{\mathrm{PL}} \propto n$, the PL intensity should scale quadratically (quartically) with the pump intensity (field strength). Indeed, the PL intensity conforms well to a power-law scaling (represented as a solid line in Fig. \ref{Fig3}a), $I_{\mathrm{PL}} \propto E^{2a}$, with $a \sim 2$, indicating a two-photon absorption process. At $\hbar\omega=0.99$ eV (see Fig. \ref{Fig3}b), we observe a similar trend with power-law scaling, indicating a three-photon process. At $\hbar\omega=0.62$ eV, while the PL intensity versus $E$ initially follows a power-law behavior at moderate field strengths, a clear deviation occurs when $E > 5$ MV/cm, marking a transition from multiphoton absorption to diabatic tunneling (Fig. \ref{Fig3}c). Lowering the photon energies to $\hbar\omega=0.50$ eV (Fig. \ref{Fig3}d) and $\hbar\omega=0.41$ eV (Figs. \ref{Fig3}d and \ref{Fig3}e) allows us to explore both the multiphoton and electron tunneling regimes by tuning $E$. At these lower pump photon energies, $\gamma$ approaches unity at a moderate field strength, but the multiphoton channel persists at lower fields. Accordingly, the PL intensity follows a power-law trend at small $E$, deviates as $E$ increases, and eventually aligns with the threshold behavior at large $E$ (i.e., $I_{\mathrm{PL}} \propto E \exp(-\pi E_{th}/E)$). As the pump photon energy is further reduced to $\hbar\omega=0.31$ eV (see Fig. \ref{Fig3}f), the PL scaling predominantly exhibits the expected threshold behavior characteristic of electron tunneling processes. This progression across different photon energies demonstrates the transition from multiphoton absorption to electron tunneling, providing direct experimental evidence for the Keldysh crossover in PL.

To rationalize our findings, we simulate the electron-hole production rate across the entire Keldysh crossover regime using the Landau-Dykhne formalism\cite{Keldysh1965IonizationWave,Kane1960ZenerSemiconductors,Oka2003BreakdownMechanism,Oka2010DielectricAnsatz,Oka2012NonlinearModel} (see Supplementary Note 4 for details). Figure \ref{Fig4}a presents the calculated rate $\Gamma$ as a function of $E$ for different pump photon energies. The simulation results clearly capture the essence of our experimental observations: a crossover from power-law scaling to threshold behavior as the Keldysh parameter is tuned from large values to below unity. Quantitatively, we can fit the simulation data to $\Gamma \propto E^{2a}$ in the multiphoton regime and to $\Gamma \propto E\exp(-\pi E_{th}/E)$ in the electron tunneling regime, allowing us to compare the number of photons $a$ participating in multiphoton absorption and the tunneling threshold $E_{th}$ with the experimentally determined values. 

In the multiphoton regime, perturbative nonlinear optics predicts that $a$ should follow a staircase function relative to $\hbar\omega$, that is, $a = [\Delta/\hbar\omega + 1]$. This is because only an integer number of photons with combined energy exceeding the interband transition energy can lead to multiphoton absorption. However, the simulated $a$ forms a smooth curve that envelops the staircase function, deviating from the expected perturbative behavior. This discrepancy arises because the Landau-Dykhne approach analytically calculates the transition rate of electron-hole production and inherently includes non-perturbative effects. Both dependencies are plotted alongside the experimentally determined $a$ in Fig. \ref{Fig4}b. Interestingly, while the experimentally determined $a$ values match the staircase function at relatively high photon energies (i.e. $a \approx 2$ at $\hbar\omega=1.55$ eV and $a \approx 3$ at $\hbar\omega=0.99$ eV), they align more closely with the simulation values at lower photon energies, where $a$ can take non-integer values. This indicates that even in the multiphoton regime ($\gamma>1$), non-perturbative effects can potentially occur at low pump photon energies, acting as a precursor to the closing of the multiphoton channel\cite{kruchinin2018colloquium}.

Similarly, in the tunneling regime, we extract the tunneling threshold $E_{th}$ at various photon energies from fits to both experimental and simulation data, as shown in Fig. \ref{Fig4}c. Using the Landau-Zener transition probability in the adiabatic tunneling regime\cite{Kane1960ZenerSemiconductors}, we obtain the tunneling threshold under a DC electric field, $E_{th}\equiv\frac{\sqrt{m\Delta^3}}{2e\hbar}=29$ MV/cm, which corresponds to the Schwinger limit\cite{schwinger1951gauge,mourou2006optics,linder2018analog}. Consequently, the electron tunneling processes induced by AC electromagnetic fields can be considered a dynamical Schwinger effect\cite{lorenc2024dynamical}. Initially, one might expect this Schwinger limit to apply at any pump photon energy within the tunneling regime. However, $E_{th}$ values derived from the simulation results seem to underestimate the AC tunneling threshold compared to the Schwinger limit, but they closely match the experimental values at intermediate pump photon energies, i.e., $\hbar\omega \sim 0.3$ eV. Notably, the experimental $E_{th}$ values deviate substantially at higher $\hbar\omega$ close to the diabatic tunneling regime. We attribute this behavior to the breakdown of the adiabatic approximation and possible multi-electron excitation processes\cite{lezius2001nonadiabatic,lezius2002polyatomic,Smits2004AbsoluteFields} (see Supplementary Note 5 for details and other possible mechanisms). As $\hbar\omega$ decreases further, the tunneling threshold seems to gradually approach the expected DC Schwinger limit. However, systematically exploring the entire low-frequency regime remains experimentally challenging since the electron-hole pair production rate is dramatically reduced with low-energy pump photons, thereby suppressing the PL\cite{zhang2023discovery,frenzel2023nonlinear}. We also note that a similar frequency-dependent tunneling threshold has been investigated in atomic gases and polyatomic molecules\cite{lezius2001nonadiabatic,lezius2002polyatomic}. To the best of our knowledge, it has never been reported in condensed matter systems. Thus, lead halide perovskites may serve as a unique solid-state platform for exploring strong-field physics.

The ability to tune the Keldysh crossover from multiphoton to electron tunneling offers a means to control the carrier distribution in reciprocal space. In the multiphoton regime, only specific portions of momentum space where the band gap matches multiples of the pump photon energy are excited. In contrast, in the tunneling regime, the entire conduction band can be populated. To explore how the carrier distribution evolves across these two regimes, we apply the Landau-Dykhne formalism to calculate the distribution of the electron-hole pair production probability across different momentum and energy states within the first Brillouin zone, as shown in Fig. \ref{Fig4}d. We first present calculations for a high pump photon energy ($\hbar\omega=1.50$ eV) at two selected pump electric fields: $E=3.75$ MV/cm and $E=25$ MV/cm, corresponding to $\gamma=9.91$ and $\gamma=1.49$, respectively. These scenarios lead to distinct charge carrier distributions over the Brillouin zone. For $\gamma=9.91$, generated carriers are mostly concentrated close to the zone center, as expected for a multiphoton process. For $\gamma=1.49$, the generated electron-hole pairs reach much larger momenta in reciprocal space, suggesting the non-perturbative occupation of conduction bands beginning in the diabatic tunneling regime. This distinction becomes more evident at a lower pump photon energy ($\hbar\omega=0.20$ eV). With a pump electric field of $E=3.75$ MV/cm ($\gamma=1.32$), electron-hole populations cluster around zero momentum, with an even denser distribution compared to that achieved with the same electric field at higher $\hbar\omega$. In contrast, when $E=25$ MV/cm, the electron tunneling process dominates ($\gamma=0.20$), resulting in an almost even distribution of generated carriers across the entire momentum space. This indicates a highly non-thermal carrier distribution. 

These results highlight the strength of using low-frequency photon fields to achieve exotic out-of-equilibrium phenomena and functionalities in semiconductors. We also envision that future research using momentum-sensitive probes, such as angle-resolved photoemission spectroscopy\cite{reimann2018subcycle,puppin2020evidence,ito2023build,boschini2024time}, will directly visualize such non-equilibrium carrier distribution functions. The tunability of the charge carrier distribution provides controlled access to specific locations within the electronic band structures, offering significant opportunities for studying the energy-momentum-dependent dynamics of electron-hole distributions. This may further open the door to generating emergent quasiparticles composed of non-zero momentum states, such as Mahan excitons\cite{mahan1967excitons,palmieri2020mahan} and Cooper-pair-like excitons\cite{keldysh1965possible,choksy2023fermi}. For practical applications, our study demonstrates the potential of using lead halide perovskites and related materials as photon upconverters and ultra-broadband sensing devices, capable of operating across the short and long-wavelength infrared regions.

\newpage
\noindent\textbf{Acknowledgments} 
We are grateful to A. Stolow, Z. Alpichshev, and T. Oka for insightful discussions. Z.Z., Z.-J.L., and K.A.N acknowledge support from the U.S. Department of Energy, Office of Basic Energy Sciences, under Award No. DE-SC0019126. H.N. and N. G. acknowledge support from the US Department of Energy, BES DMSE. E.B. was primarily supported by the Robert A. Welch Foundation under grant F-2092-20220331. A.D.M. acknowledges support from the Army Research Office under grant W911NF2210158. J.H. acknowledges financial support from the China Scholarships Council (number 202107990007). 

\noindent\textbf{Author contributions}
Z.Z. designed the project. Z.Z. and Z.-J. L. built the experimental setup and performed all the measurements. Z.Z. and H. N. analyzed the data. H. N. performed theoretical calculations under the supervision of N. G. J. H. synthesized and characterized CsPbBr$_3$ single crystals under the supervision of A. D. M. Z.Z., H. N., and Z.-J. L. led the manuscript preparation with crucial input from E.B. and K.A.N. and contributions from all authors. K.A.N. supervised the project. 

\noindent\textbf{Competing interests} The authors declare no competing interests.

\noindent\textbf{Data availability}
Source data are provided with this paper. All other data that support the findings of this study are available from the corresponding author upon reasonable request.

\noindent\textbf{Code availability}
The codes used to perform the simulations and to analyze the data in this work are available from the corresponding author upon reasonable request.

\section*{Methods}
\noindent \textit{Synthesis of CsPbBr$_3$ single crystals}\\
PbO (223 mg, 1 mmol) was dissolved in HBr (3 ml) at room temperature on a hotplate with vigorous stirring. Then CsBr (212 mg, 1 mmol) was added to this solution, forming a bright orange powder (CsPbBr$_3$). The temperature of the hotplate was increased to 150 $^\circ$C with continued stirring to dissolve the orange powder completely. The temperature was then lowered to 25 $^\circ$C, resulting in the formation of plate-shaped CsPbBr$_3$ seed crystals ($\sim$0.1 mm lateral size).

The growth of thin CsPbBr$_3$ single crystals was achieved by combining two strategies: slow crystallization (by slow cooling of a dilute HBr solution\cite{liu2018recent}) and thickness control (by space confinement\cite{hou2024synthesis}). In a separate beaker, PbO (112 mg, 0.5 mmol) was dissolved in HBr (3.5 ml) at room temperature on a hotplate with vigorous stirring. Then CsBr (106 mg, 0.5 mmol) was added to this solution, forming a bright orange powder (CsPbBr$_3$), which soon completely dissolved. The temperature of the hotplate was increased to 150 $^\circ$C, and stirring was stopped.

A sapphire wafer (10 mm $\times$ 10 mm $\times$ 1 mm, two sides polished, MTI Corp.) was used as the substrate for space confinement. Sapphire wafers were cleaned in soapy water, acetone, and isopropanol by ultrasonication for 20 minutes each, then dried with argon. The sapphire wafers were transferred into a UV-ozone cleaner and cleaned for 15 minutes. A CsPbBr$_3$ seed crystal was placed between two sapphire wafers, and this stack was placed at the bottom of the hot CsPbBr$_3$ solution mentioned above. The temperature of the hotplate was slowly cooled to 35 $^\circ$C (at a rate of 1 $^\circ$C per 4 seconds) and maintained for 1 week. The CsPbBr$_3$ seed crystal slowly grew into a larger thin crystal ($\sim$2 mm lateral size). The crystal was filtered and washed with heptane.

\noindent \textit{Experimental details}\\
Our experimental setup consists of the pump source, the sample, and the PL detection system. To generate pump pulses at different photon energies throughout the infrared frequency range ($\hbar\omega=0.25$--$1.55$ eV), we adopted various frequency-conversion approaches. In the near-infrared regime (0.50 eV $\leq\hbar\omega\leq$ 1.55 eV), the pump pulses were either derived from a portion of the fundamental laser output (i.e., $\hbar\omega=1.55$ eV, 800 nm) from a Ti:Sapphire regenerative amplifier (Coherent Legend Elite Duo, 13 mJ, 1 kHz), or from the signal and idler beams of a high-energy optical parametric amplifier (OPA, Light Conversion TOPAS-Prime-HE) seeded by the Ti:Sapphire amplifier. In the mid-infrared regime ($\hbar\omega\leq$ 0.50 eV), pump pulses were generated by difference frequency mixing of the signal and idler outputs in a 0.5-mm thick GaSe crystal. The pulse duration of all pump pulses was slightly stretched and compensated to be approximately 200 fs. The spectral characterizations of the pump pulses were achieved by either directly measuring the spectrum or employing nonlinear frequency upconversion methods (see Supplementary Note 1 for details).

The generated pump pulses were collected by a parabolic mirror with a 3-inch reflective focal length and then focused onto the sample, with a beam diameter of 100--350 $\mu$m, depending on the exact pump wavelength. The sample was kept inside a vacuum cell with calcium fluoride windows. The PL emission from the sample was collected by a 50x objective (Mitutoyo) and detected by a spectrometer (Andor Kymera 193i) or a photomultiplier tube (Hamamatsu H10720).

For electric field-dependent measurements, two wire-grid polarizers (Thorlab WP25H-Z) were used to control the pump electric field strength in the mid-infrared frequency range, while for near-infrared pulses, a half-wave plate and a thin-film polarizer were used for electric field control.

\noindent \textit{Fitting procedure}\\
The PL intensity ($I_{\mathrm{PL}}$) as a function of the pump electric field strength ($E$) in the multiphoton regime was fit with $I(E)\propto E^{2a}$. The upper bound of the fitting range was determined as follows: we first plotted $I(E)$ on a bi-logarithmic scale. A deviation from linearity was observed for each pump energy when $\gamma<2\mbox{--}3$. We thus fit the linear region with the power-law formula. Similarly, in the adiabatic tunneling regime, with a lower bound determined by $\gamma\lesssim1$, the data were fit with $I(E)\propto E\exp(-\pi E_{th}/E)$. We extended the lower bound to $\gamma$ slightly larger than 1 to include more data points in the fitting for several photon energies. The variation in the fitted $E_{th}$ value does not exceed the fitting error bar. A detailed derivation of the fitting formulas is discussed in Supplementary Note 4.

\noindent \textit{Simulation details}\\
We used the Landau-Dykhne formalism to simulate the electron-hole pair production rate $\Gamma$ in CsPbBr$_3$. This method has been shown to effectively capture the entire Keldysh crossover regime, from multiphoton excitation to quantum electron tunneling in atomic, semiconducting, and correlated insulator systems \cite{Keldysh1965IonizationWave,Kane1960ZenerSemiconductors,Oka2010DielectricAnsatz,Oka2012NonlinearModel}. We assume a one-dimensional two-band semiconductor model and calculate the momentum-dependent transition probability $P_p$ based on the adiabatic perturbative theory. The total electron-hole pair production rate $\Gamma$ can then be calculated by integrating $P_p$ over the momentum space as a function of $\hbar\omega$ and $E$. A more detailed discussion can be found in Supplementary Note 4.

\newpage

\newpage

\newpage

\newpage
\begin{center}
\begin{figure}[H]
\linespread{1}
   \sbox0{\includegraphics[width=\textwidth]{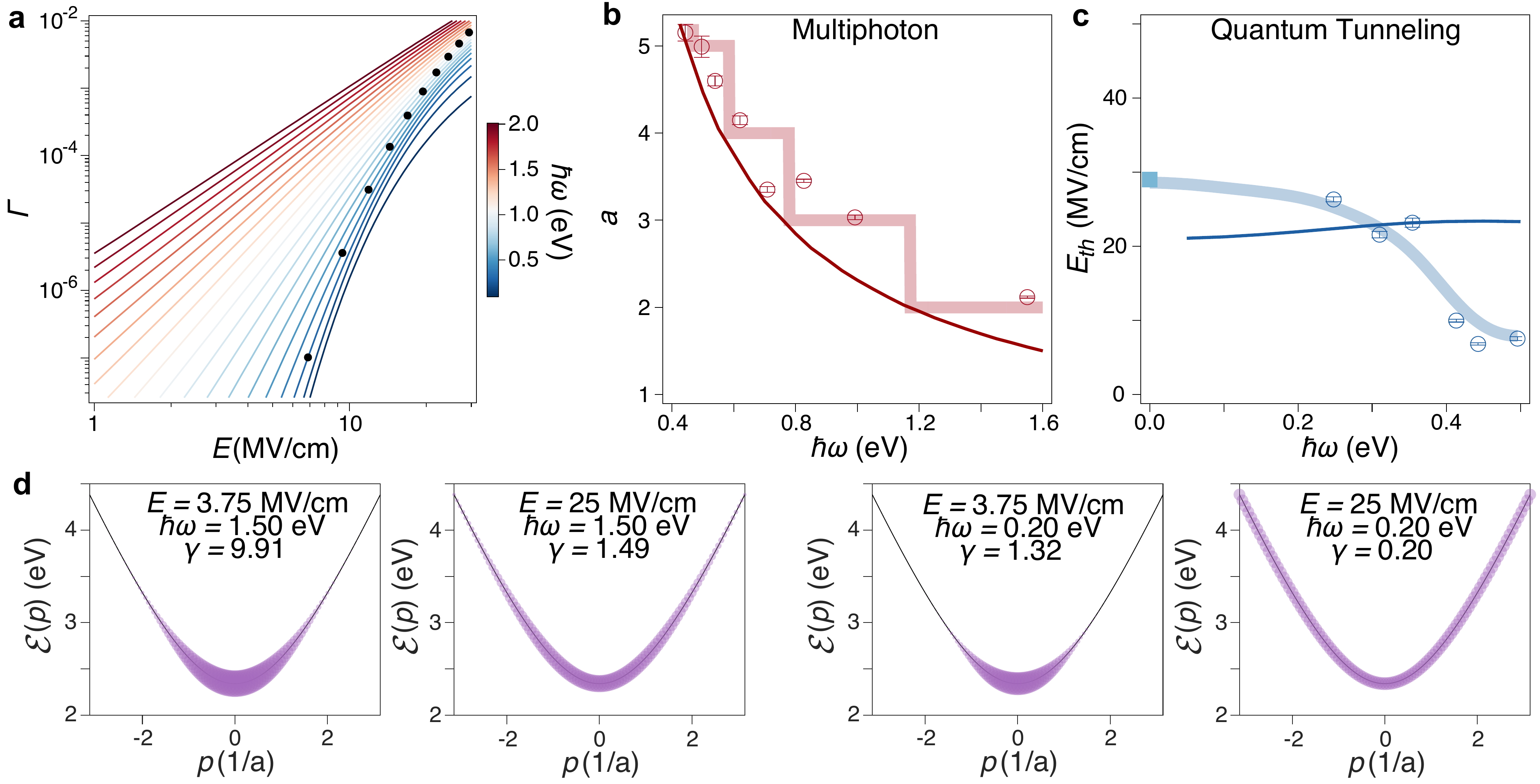}}% measure width
    \begin{minipage}{\wd0}
  \usebox0
  %\captionsetup{justification=raggedright,singlelinecheck=false}
  \captionsetup{labelfont={bf},labelformat={default},labelsep=period,name={Fig.}}
  \caption{\textbf{Simulation results and comparison to experimental fitting parameters.} \textbf{a,} Landau-Dykhne-formalism-based simulation of the electron-hole pair production rate $\Gamma$ as a function of the driving field strength $E$ at a wide range of pump photon energies $\hbar\omega$. Black dots indicate where $\gamma=1$ is realized. \textbf{b,} Multiphoton number $a$ as a function of the pump photon energy $\hbar\omega$ obtained from fits to the experimental data shown in Fig.3 (red circles) and from fits to the simulation results in panel \textbf{a} (thin solid line). The thick line represents the prediction under the perturbative multiphoton assumption: $[\Delta/\hbar\omega+1]$, where the square brackets denote the floor function and $\Delta$ is the bandgap. \textbf{c,} Tunneling threshold $E_{th}$ as a function of the pump photon energy $\hbar\omega$, obtained from fits to the experimental data shown in Fig.3 (blue circles) and from fits to the simulation results shown in panel \textbf{a} (thin solid line). The light blue square marks the value of $E_{th}=29$ MV/cm in the DC Schwinger limit, and the thick line serves as a guide for the eye. \textbf{d,} Simulated energy-momentum dependent electron-hole pair production probability at various $\hbar\omega$ and $E$ values across the Keldysh crossover regime. $\mathcal{E}(q)$ represents the total energy of a single electron-hole pair at a central momentum $q$. The size of the circles characterizes the probability value at each momentum in the first Brillouin zone. Circles in the lower $E$ panels are enlarged by $500$ and $10^8$ times compared to the higher $E$ panels for 1.50 eV and 0.20 eV, respectively. The error bars in panels \textbf{b} and \textbf{c} are obtained from the standard deviation of the corresponding fitting results.} 
  \label{Fig4}
\end{minipage}
\end{figure}  
\end{center}

\newpage
\section*{References}
\footnotesize
\bibliographystyle{naturemag}
\bibliography{References}

\begin{thebibliography}{10}
\expandafter\ifx\csname url\endcsname\relax
  \def\url#1{\texttt{#1}}\fi
\expandafter\ifx\csname urlprefix\endcsname\relax\def\urlprefix{URL }\fi
\providecommand{\bibinfo}[2]{#2}
\providecommand{\eprint}[2][]{\url{#2}}

\bibitem{sutherland2016perovskite}
\bibinfo{author}{Sutherland, B.~R.} \& \bibinfo{author}{Sargent, E.~H.}
\newblock \bibinfo{title}{Perovskite photonic sources}.
\newblock \emph{\bibinfo{journal}{Nature Photonics}}
  \textbf{\bibinfo{volume}{10}}, \bibinfo{pages}{295--302}
  (\bibinfo{year}{2016}).

\bibitem{kovalenko2017properties}
\bibinfo{author}{Kovalenko, M.~V.}, \bibinfo{author}{Protesescu, L.} \&
  \bibinfo{author}{Bodnarchuk, M.~I.}
\newblock \bibinfo{title}{Properties and potential optoelectronic applications
  of lead halide perovskite nanocrystals}.
\newblock \emph{\bibinfo{journal}{Science}} \textbf{\bibinfo{volume}{358}},
  \bibinfo{pages}{745--750} (\bibinfo{year}{2017}).

\bibitem{fu2019metal}
\bibinfo{author}{Fu, Y.} \emph{et~al.}
\newblock \bibinfo{title}{Metal halide perovskite nanostructures for
  optoelectronic applications and the study of physical properties}.
\newblock \emph{\bibinfo{journal}{Nature Reviews Materials}}
  \textbf{\bibinfo{volume}{4}}, \bibinfo{pages}{169--188}
  (\bibinfo{year}{2019}).

\bibitem{zhao2016organic}
\bibinfo{author}{Zhao, Y.} \& \bibinfo{author}{Zhu, K.}
\newblock \bibinfo{title}{Organic--inorganic hybrid lead halide perovskites for
  optoelectronic and electronic applications}.
\newblock \emph{\bibinfo{journal}{Chemical Society Reviews}}
  \textbf{\bibinfo{volume}{45}}, \bibinfo{pages}{655--689}
  (\bibinfo{year}{2016}).

\bibitem{younis2021halide}
\bibinfo{author}{Younis, A.} \emph{et~al.}
\newblock \bibinfo{title}{Halide perovskites: a new era of solution-processed
  electronics}.
\newblock \emph{\bibinfo{journal}{Advanced Materials}}
  \textbf{\bibinfo{volume}{33}}, \bibinfo{pages}{2005000}
  (\bibinfo{year}{2021}).

\bibitem{chen2022two}
\bibinfo{author}{Chen, J.}, \bibinfo{author}{Zhang, W.} \&
  \bibinfo{author}{Pullerits, T.}
\newblock \bibinfo{title}{Two-photon absorption in halide perovskites and their
  applications}.
\newblock \emph{\bibinfo{journal}{Materials Horizons}}
  \textbf{\bibinfo{volume}{9}}, \bibinfo{pages}{2255--2287}
  (\bibinfo{year}{2022}).

\bibitem{zhou2021perovskites}
\bibinfo{author}{Zhou, F.}, \bibinfo{author}{Ran, X.}, \bibinfo{author}{Fan,
  D.}, \bibinfo{author}{Lu, S.} \& \bibinfo{author}{Ji, W.}
\newblock \bibinfo{title}{Perovskites: multiphoton absorption and
  applications}.
\newblock \emph{\bibinfo{journal}{Advanced Optical Materials}}
  \textbf{\bibinfo{volume}{9}}, \bibinfo{pages}{2100292}
  (\bibinfo{year}{2021}).

\bibitem{fakharuddin2022perovskite}
\bibinfo{author}{Fakharuddin, A.} \emph{et~al.}
\newblock \bibinfo{title}{Perovskite light-emitting diodes}.
\newblock \emph{\bibinfo{journal}{Nature Electronics}}
  \textbf{\bibinfo{volume}{5}}, \bibinfo{pages}{203--216}
  (\bibinfo{year}{2022}).

\bibitem{liu2021electroluminescence}
\bibinfo{author}{Liu, A.} \emph{et~al.}
\newblock \bibinfo{title}{Electroluminescence principle and performance
  improvement of metal halide perovskite light-emitting diodes}.
\newblock \emph{\bibinfo{journal}{Advanced Optical Materials}}
  \textbf{\bibinfo{volume}{9}}, \bibinfo{pages}{2002167}
  (\bibinfo{year}{2021}).

\bibitem{kruchinin2018colloquium}
\bibinfo{author}{Kruchinin, S.~Y.}, \bibinfo{author}{Krausz, F.} \&
  \bibinfo{author}{Yakovlev, V.~S.}
\newblock \bibinfo{title}{Colloquium: Strong-field phenomena in periodic
  systems}.
\newblock \emph{\bibinfo{journal}{Reviews of Modern Physics}}
  \textbf{\bibinfo{volume}{90}}, \bibinfo{pages}{021002}
  (\bibinfo{year}{2018}).

\bibitem{kuehn2010terahertz}
\bibinfo{author}{Kuehn, W.} \emph{et~al.}
\newblock \bibinfo{title}{Terahertz-induced interband tunneling of electrons in
  gaas}.
\newblock \emph{\bibinfo{journal}{Physical Review B}}
  \textbf{\bibinfo{volume}{82}}, \bibinfo{pages}{075204}
  (\bibinfo{year}{2010}).

\bibitem{hirori2011extraordinary}
\bibinfo{author}{Hirori, H.} \emph{et~al.}
\newblock \bibinfo{title}{Extraordinary carrier multiplication gated by a
  picosecond electric field pulse}.
\newblock \emph{\bibinfo{journal}{Nature Communications}}
  \textbf{\bibinfo{volume}{2}}, \bibinfo{pages}{594} (\bibinfo{year}{2011}).

\bibitem{Ghimire2011RedshiftField}
\bibinfo{author}{Ghimire, S.} \emph{et~al.}
\newblock \bibinfo{title}{{Redshift in the optical absorption of ZnO single
  crystals in the presence of an intense midinfrared laser field}}.
\newblock \emph{\bibinfo{journal}{Physical Review Letters}}
  \textbf{\bibinfo{volume}{107}}, \bibinfo{pages}{167407}
  (\bibinfo{year}{2011}).

\bibitem{schultze2014attosecond}
\bibinfo{author}{Schultze, M.} \emph{et~al.}
\newblock \bibinfo{title}{Attosecond band-gap dynamics in silicon}.
\newblock \emph{\bibinfo{journal}{Science}} \textbf{\bibinfo{volume}{346}},
  \bibinfo{pages}{1348--1352} (\bibinfo{year}{2014}).

\bibitem{lange2014extremely}
\bibinfo{author}{Lange, C.} \emph{et~al.}
\newblock \bibinfo{title}{Extremely nonperturbative nonlinearities in {GaAs}
  driven by atomically strong terahertz fields in gold metamaterials}.
\newblock \emph{\bibinfo{journal}{Physical Review Letters}}
  \textbf{\bibinfo{volume}{113}}, \bibinfo{pages}{227401}
  (\bibinfo{year}{2014}).

\bibitem{Mayer2015TunnelingExcitation}
\bibinfo{author}{Mayer, B.} \emph{et~al.}
\newblock \bibinfo{title}{{Tunneling breakdown of a strongly correlated
  insulating state in {VO$_2$} induced by intense multiterahertz excitation}}.
\newblock \emph{\bibinfo{journal}{Physical Review B}}
  \textbf{\bibinfo{volume}{91}}, \bibinfo{pages}{235113}
  (\bibinfo{year}{2015}).

\bibitem{pein2017terahertz}
\bibinfo{author}{Pein, B.~C.} \emph{et~al.}
\newblock \bibinfo{title}{Terahertz-driven luminescence and colossal {stark}
  effect in {CdSe--CdS} colloidal quantum dots}.
\newblock \emph{\bibinfo{journal}{Nano Letters}} \textbf{\bibinfo{volume}{17}},
  \bibinfo{pages}{5375--5380} (\bibinfo{year}{2017}).

\bibitem{shi2022terahertz}
\bibinfo{author}{Shi, J.} \emph{et~al.}
\newblock \bibinfo{title}{Terahertz field-induced reemergence of quenched
  photoluminescence in quantum dots}.
\newblock \emph{\bibinfo{journal}{Nano Letters}} \textbf{\bibinfo{volume}{22}},
  \bibinfo{pages}{1718--1725} (\bibinfo{year}{2022}).

\bibitem{li2022keldysh}
\bibinfo{author}{Li, X.} \emph{et~al.}
\newblock \bibinfo{title}{{Keldysh} space control of charge dynamics in a
  strongly driven {Mott} insulator}.
\newblock \emph{\bibinfo{journal}{Physical Review Letters}}
  \textbf{\bibinfo{volume}{128}}, \bibinfo{pages}{187402}
  (\bibinfo{year}{2022}).

\bibitem{nishidome2023influence}
\bibinfo{author}{Nishidome, H.} \emph{et~al.}
\newblock \bibinfo{title}{Influence of laser intensity and location of the
  {Fermi} level on tunneling processes for high-harmonic generation in arrayed
  semiconducting carbon nanotubes}.
\newblock \emph{\bibinfo{journal}{ACS Photonics}}
  \textbf{\bibinfo{volume}{11}}, \bibinfo{pages}{171--179}
  (\bibinfo{year}{2023}).

\bibitem{shinjo2024keldysh}
\bibinfo{author}{Shinjo, K.} \& \bibinfo{author}{Tohyama, T.}
\newblock \bibinfo{title}{{Keldysh} crossover in one-dimensional {Mott}
  insulators}.
\newblock \emph{\bibinfo{journal}{APL Materials}} \textbf{\bibinfo{volume}{12}}
  (\bibinfo{year}{2024}).

\bibitem{Keldysh1965IonizationWave}
\bibinfo{author}{Keldysh, L.~V.}
\newblock \bibinfo{title}{{Ionization in the Field of a Strong Electromagnetic
  Wave}}.
\newblock In \emph{\bibinfo{booktitle}{Soviet Physics - JETP}},
  vol.~\bibinfo{volume}{20}, \bibinfo{pages}{1307--1314}
  (\bibinfo{year}{1965}).

\bibitem{saouma2017multiphoton}
\bibinfo{author}{Saouma, F.~O.}, \bibinfo{author}{Stoumpos, C.~C.},
  \bibinfo{author}{Kanatzidis, M.~G.}, \bibinfo{author}{Kim, Y.~S.} \&
  \bibinfo{author}{Jang, J.~I.}
\newblock \bibinfo{title}{Multiphoton absorption order of {CsPbBr$_3$} as
  determined by wavelength-dependent nonlinear optical spectroscopy}.
\newblock \emph{\bibinfo{journal}{The Journal of Physical Chemistry Letters}}
  \textbf{\bibinfo{volume}{8}}, \bibinfo{pages}{4912--4917}
  (\bibinfo{year}{2017}).

\bibitem{clark2016polarization}
\bibinfo{author}{Clark, D.}, \bibinfo{author}{Stoumpos, C.},
  \bibinfo{author}{Saouma, F.}, \bibinfo{author}{Kanatzidis, M.} \&
  \bibinfo{author}{Jang, J.}
\newblock \bibinfo{title}{Polarization-selective three-photon absorption and
  subsequent photoluminescence in {CsPbBr$_3$} single crystal at room
  temperature}.
\newblock \emph{\bibinfo{journal}{Physical Review B}}
  \textbf{\bibinfo{volume}{93}}, \bibinfo{pages}{195202}
  (\bibinfo{year}{2016}).

\bibitem{abiedh2021mixed}
\bibinfo{author}{Abiedh, K.}, \bibinfo{author}{Zaaboub, Z.} \&
  \bibinfo{author}{Hassen, F.}
\newblock \bibinfo{title}{{Mixed monomolecular and bimolecular-like
  recombination processes in {CsPbBr$_3$} perovskite film revealed by
  time-resolved photoluminescence spectroscopy}}.
\newblock \emph{\bibinfo{journal}{Applied Physics A}}
  \textbf{\bibinfo{volume}{127}}, \bibinfo{pages}{1--9} (\bibinfo{year}{2021}).

\bibitem{Kane1960ZenerSemiconductors}
\bibinfo{author}{Kane, E.~O.}
\newblock \bibinfo{title}{{{Zener} tunneling in semiconductors}}.
\newblock \emph{\bibinfo{journal}{Journal of Physics and Chemistry of Solids}}
  \textbf{\bibinfo{volume}{12}}, \bibinfo{pages}{181--188}
  (\bibinfo{year}{1960}).

\bibitem{Oka2003BreakdownMechanism}
\bibinfo{author}{Oka, T.}, \bibinfo{author}{Arita, R.} \&
  \bibinfo{author}{Aoki, H.}
\newblock \bibinfo{title}{{Breakdown of a {Mott} Insulator: A Nonadiabatic
  Tunneling Mechanism}}.
\newblock \emph{\bibinfo{journal}{Physical Review Letters}}
  \textbf{\bibinfo{volume}{91}}, \bibinfo{pages}{066406}
  (\bibinfo{year}{2003}).

\bibitem{Oka2010DielectricAnsatz}
\bibinfo{author}{Oka, T.} \& \bibinfo{author}{Aoki, H.}
\newblock \bibinfo{title}{{Dielectric breakdown in a {Mott} insulator:
  Many-body {Schwinger-Landau-Zener} mechanism studied with a generalized Bethe
  ansatz}}.
\newblock \emph{\bibinfo{journal}{Physical Review B}}
  \textbf{\bibinfo{volume}{81}}, \bibinfo{pages}{033103}
  (\bibinfo{year}{2010}).

\bibitem{Oka2012NonlinearModel}
\bibinfo{author}{Oka, T.}
\newblock \bibinfo{title}{{Nonlinear doublon production in a {Mott} insulator:
  {Landau-Dykhne} method applied to an integrable model}}.
\newblock \emph{\bibinfo{journal}{Physical Review B}}
  \textbf{\bibinfo{volume}{86}}, \bibinfo{pages}{075148}
  (\bibinfo{year}{2012}).

\bibitem{schwinger1951gauge}
\bibinfo{author}{Schwinger, J.}
\newblock \bibinfo{title}{On gauge invariance and vacuum polarization}.
\newblock \emph{\bibinfo{journal}{Physical Review}}
  \textbf{\bibinfo{volume}{82}}, \bibinfo{pages}{664} (\bibinfo{year}{1951}).

\bibitem{mourou2006optics}
\bibinfo{author}{Mourou, G.~A.}, \bibinfo{author}{Tajima, T.} \&
  \bibinfo{author}{Bulanov, S.~V.}
\newblock \bibinfo{title}{Optics in the relativistic regime}.
\newblock \emph{\bibinfo{journal}{Reviews of Modern Physics}}
  \textbf{\bibinfo{volume}{78}}, \bibinfo{pages}{309--371}
  (\bibinfo{year}{2006}).

\bibitem{linder2018analog}
\bibinfo{author}{Linder, M.~F.}, \bibinfo{author}{Lorke, A.} \&
  \bibinfo{author}{Sch{\"u}tzhold, R.}
\newblock \bibinfo{title}{Analog {Sauter-Schwinger} effect in semiconductors
  for spacetime-dependent fields}.
\newblock \emph{\bibinfo{journal}{Physical Review B}}
  \textbf{\bibinfo{volume}{97}}, \bibinfo{pages}{035203}
  (\bibinfo{year}{2018}).

\bibitem{lorenc2024dynamical}
\bibinfo{author}{Lorenc, D.} \emph{et~al.}
\newblock \bibinfo{title}{Dynamical schwinger effect in lead-halide
  perovskites}.
\newblock \emph{\bibinfo{journal}{arXiv preprint arXiv:2406.05032}}
  (\bibinfo{year}{2024}).

\bibitem{lezius2001nonadiabatic}
\bibinfo{author}{Lezius, M.} \emph{et~al.}
\newblock \bibinfo{title}{Nonadiabatic multielectron dynamics in strong field
  molecular ionization}.
\newblock \emph{\bibinfo{journal}{Physical Review Letters}}
  \textbf{\bibinfo{volume}{86}}, \bibinfo{pages}{51} (\bibinfo{year}{2001}).

\bibitem{lezius2002polyatomic}
\bibinfo{author}{Lezius, M.}, \bibinfo{author}{Blanchet, V.},
  \bibinfo{author}{Ivanov, M.~Y.} \& \bibinfo{author}{Stolow, A.}
\newblock \bibinfo{title}{Polyatomic molecules in strong laser fields:
  Nonadiabatic multielectron dynamics}.
\newblock \emph{\bibinfo{journal}{The Journal of Chemical Physics}}
  \textbf{\bibinfo{volume}{117}}, \bibinfo{pages}{1575--1588}
  (\bibinfo{year}{2002}).

\bibitem{Smits2004AbsoluteFields}
\bibinfo{author}{Smits, M.}, \bibinfo{author}{De~Lange, C.~A.},
  \bibinfo{author}{Stolow, A.} \& \bibinfo{author}{Rayner, D.~M.}
\newblock \bibinfo{title}{{Absolute ionization rates of multielectron
  transition metal atoms in strong infrared laser fields}}.
\newblock \emph{\bibinfo{journal}{Physical Review Letters}}
  \textbf{\bibinfo{volume}{93}}, \bibinfo{pages}{213003}
  (\bibinfo{year}{2004}).

\bibitem{zhang2023discovery}
\bibinfo{author}{Zhang, Z.} \emph{et~al.}
\newblock \bibinfo{title}{Discovery of enhanced lattice dynamics in a
  single-layered hybrid perovskite}.
\newblock \emph{\bibinfo{journal}{Science Advances}}
  \textbf{\bibinfo{volume}{9}}, \bibinfo{pages}{eadg4417}
  (\bibinfo{year}{2023}).

\bibitem{frenzel2023nonlinear}
\bibinfo{author}{Frenzel, M.} \emph{et~al.}
\newblock \bibinfo{title}{Nonlinear terahertz control of the lead halide
  perovskite lattice}.
\newblock \emph{\bibinfo{journal}{Science Advances}}
  \textbf{\bibinfo{volume}{9}}, \bibinfo{pages}{eadg3856}
  (\bibinfo{year}{2023}).

\bibitem{reimann2018subcycle}
\bibinfo{author}{Reimann, J.} \emph{et~al.}
\newblock \bibinfo{title}{Subcycle observation of lightwave-driven dirac
  currents in a topological surface band}.
\newblock \emph{\bibinfo{journal}{Nature}} \textbf{\bibinfo{volume}{562}},
  \bibinfo{pages}{396--400} (\bibinfo{year}{2018}).

\bibitem{puppin2020evidence}
\bibinfo{author}{Puppin, M.} \emph{et~al.}
\newblock \bibinfo{title}{Evidence of large polarons in photoemission band
  mapping of the perovskite semiconductor {CsPbBr$_3$}}.
\newblock \emph{\bibinfo{journal}{Physical review letters}}
  \textbf{\bibinfo{volume}{124}}, \bibinfo{pages}{206402}
  (\bibinfo{year}{2020}).

\bibitem{ito2023build}
\bibinfo{author}{Ito, S.} \emph{et~al.}
\newblock \bibinfo{title}{Build-up and dephasing of {Floquet--Bloch} bands on
  subcycle timescales}.
\newblock \emph{\bibinfo{journal}{Nature}} \textbf{\bibinfo{volume}{616}},
  \bibinfo{pages}{696--701} (\bibinfo{year}{2023}).

\bibitem{boschini2024time}
\bibinfo{author}{Boschini, F.}, \bibinfo{author}{Zonno, M.} \&
  \bibinfo{author}{Damascelli, A.}
\newblock \bibinfo{title}{Time-resolved {ARPES} studies of quantum materials}.
\newblock \emph{\bibinfo{journal}{Reviews of Modern Physics}}
  \textbf{\bibinfo{volume}{96}}, \bibinfo{pages}{015003}
  (\bibinfo{year}{2024}).

\bibitem{mahan1967excitons}
\bibinfo{author}{Mahan, G.}
\newblock \bibinfo{title}{Excitons in degenerate semiconductors}.
\newblock \emph{\bibinfo{journal}{Physical Review}}
  \textbf{\bibinfo{volume}{153}}, \bibinfo{pages}{882} (\bibinfo{year}{1967}).

\bibitem{palmieri2020mahan}
\bibinfo{author}{Palmieri, T.} \emph{et~al.}
\newblock \bibinfo{title}{Mahan excitons in room-temperature methylammonium
  lead bromide perovskites}.
\newblock \emph{\bibinfo{journal}{Nature Communications}}
  \textbf{\bibinfo{volume}{11}}, \bibinfo{pages}{850} (\bibinfo{year}{2020}).

\bibitem{keldysh1965possible}
\bibinfo{author}{Keldysh, L.}
\newblock \bibinfo{title}{Possible instability of semimetallic state towards
  coulomb interaction}.
\newblock \emph{\bibinfo{journal}{Sov. Phys. Solid State}}
  \textbf{\bibinfo{volume}{6}}, \bibinfo{pages}{2219} (\bibinfo{year}{1965}).

\bibitem{choksy2023fermi}
\bibinfo{author}{Choksy, D.}, \bibinfo{author}{Szwed, E.},
  \bibinfo{author}{Butov, L.}, \bibinfo{author}{Baldwin, K.} \&
  \bibinfo{author}{Pfeiffer, L.}
\newblock \bibinfo{title}{Fermi edge singularity in neutral electron--hole
  system}.
\newblock \emph{\bibinfo{journal}{Nature Physics}}
  \textbf{\bibinfo{volume}{19}}, \bibinfo{pages}{1275--1279}
  (\bibinfo{year}{2023}).

\bibitem{liu2018recent}
\bibinfo{author}{Liu, Y.}, \bibinfo{author}{Yang, Z.} \& \bibinfo{author}{Liu,
  S.}
\newblock \bibinfo{title}{Recent progress in single-crystalline perovskite
  research including crystal preparation, property evaluation, and
  applications}.
\newblock \emph{\bibinfo{journal}{Advanced Science}}
  \textbf{\bibinfo{volume}{5}}, \bibinfo{pages}{1700471}
  (\bibinfo{year}{2018}).

\bibitem{hou2024synthesis}
\bibinfo{author}{Hou, J.} \emph{et~al.}
\newblock \bibinfo{title}{Synthesis of {2D} perovskite crystals via progressive
  transformation of quantum well thickness}.
\newblock \emph{\bibinfo{journal}{Nature Synthesis}}
  \textbf{\bibinfo{volume}{3}}, \bibinfo{pages}{265--275}
  (\bibinfo{year}{2024}).

\end{thebibliography}
\end{document}